\documentclass[aps,twocolumn,superscriptaddress,prx]{revtex4-2}
\usepackage[utf8]{inputenc}
\usepackage{amsmath}
\usepackage{amssymb}
\usepackage{graphicx}
\usepackage{color}
\usepackage{varwidth}
\usepackage[bookmarks=false]{hyperref}
\usepackage[usenames,dvipsnames]{xcolor}
\usepackage[super]{nth}
\usepackage{ulem,lipsum}
\usepackage{subcaption} 
\usepackage[export]{adjustbox} 
\usepackage{cleveref} 
\usepackage{tikz}
\usetikzlibrary{shapes}
\def\be{\begin{equation}}
\def\ee{\end{equation}}
\def\ba{\begin{eqnarray}}
\def\ea{\end{eqnarray}}

\begin{document}

\title{Possible Unconventional Surface Superconductivity in the Half-Heusler YPtBi}

\author{Eylon Persky}
\email{perskye1@stanford.edu}
\affiliation{Geballe Laboratory for Advanced Materials, Stanford University, Stanford, CA 94305, USA}
\affiliation{Stanford Institute for Materials and Energy Sciences, SLAC National Accelerator Laboratory, 2575 Sand Hill Road, Menlo Park, CA 94025, USA}
\affiliation{Department of Applied Physics, Stanford University, Stanford, CA 94305, USA}
\author{Alan Fang}
\affiliation{Geballe Laboratory for Advanced Materials, Stanford University, Stanford, CA 94305, USA}
\affiliation{Stanford Institute for Materials and Energy Sciences, SLAC National Accelerator Laboratory, 2575 Sand Hill Road, Menlo Park, CA 94025, USA}
\author{Xinyang Zhang}
\affiliation{Geballe Laboratory for Advanced Materials, Stanford University, Stanford, CA 94305, USA}
\affiliation{Stanford Institute for Materials and Energy Sciences, SLAC National Accelerator Laboratory, 2575 Sand Hill Road, Menlo Park, CA 94025, USA}
\affiliation{Department of Applied Physics, Stanford University, Stanford, CA 94305, USA}
\author{Carolina Adamo}
\altaffiliation[Current address: ]{Northrop Grumman Corporation, Redondo Beach, CA 90278, USA}
\affiliation{Geballe Laboratory for Advanced Materials, Stanford University, Stanford, CA 94305, USA}
\affiliation{Stanford Institute for Materials and Energy Sciences, SLAC National Accelerator Laboratory, 2575 Sand Hill Road, Menlo Park, CA 94025, USA}
\author{Phillip Wu}
\altaffiliation[Current address: ]{Center of Science Education, National Chung-Hsing University, Department of Chemistry, Taichung City 402, Taiwan ROC}
\affiliation{Geballe Laboratory for Advanced Materials, Stanford University, Stanford, CA 94305, USA}
\author{Eli Levenson-Falk}
\altaffiliation[Current address: ]{Center for Quantum Information Science and Technology, University of Southern California, Los Angeles, California 90089, USA}
\affiliation{Geballe Laboratory for Advanced Materials, Stanford University, Stanford, CA 94305, USA}
\affiliation{Department of Applied Physics, Stanford University, Stanford, CA 94305, USA}
\author{Chandra Shekhar}
\affiliation{Max Planck Institute for Chemical Physics of Solids, 01187 Dresden, Germany.}
\author{Claudia Felser}
\affiliation{Max Planck Institute for Chemical Physics of Solids, 01187 Dresden, Germany.}
\author{Binghai Yan}
\affiliation{Department of Condensed Matter Physics, Weizmann Institute of Science, Rehovot, Israel}
\author{Aharon Kapitulnik}
\email{aharonk@stanford.edu}
\affiliation{Geballe Laboratory for Advanced Materials, Stanford University, Stanford, CA 94305, USA}
\affiliation{Stanford Institute for Materials and Energy Sciences, SLAC National Accelerator Laboratory, 2575 Sand Hill Road, Menlo Park, CA 94025, USA}
\affiliation{Department of Applied Physics, Stanford University, Stanford, CA 94305, USA}
\affiliation{Department of Physics, Stanford University, Stanford, CA 94305, USA}

\begin{abstract}
We report an extensive study of the noncentrosymmetric half-Heusler superconductor YPtBi, revealing an unusual relation between bulk superconductivity and the possible appearance of surface superconductivity on the (111) oriented surface, at temperatures up to 3 times the bulk transition temperature. Transport measurements confirmed the low carrier density of the material and its bulk superconducting transition, which was also observed in ac susceptibility through mutual inductance (MI) measurements. However, a weak signature of superconductivity in the MI measurements appeared much above the bulk transition temperature, which was further observed in scanning tunneling spectroscopy, pointing to a possible surface superconducting state. Polar Kerr effect measurements suggest that while the bulk superconductor may exhibit an unusual nodal superconducting state, only the surface state breaks time reversal symmetry. Complementary tunneling measurements on LuPtBi are used to establish the observations on YPtBi, while density-functional theory calculations may shed light on the origin of this unusual surface state.

\end{abstract}

\date{\today}

\maketitle

Topological superconductivity can be realized by inducing superconductivity at the surface of a three dimensional topological insulator \cite{Fu2008}. This can be achieved through interface engineering, or by using the surface states of a bulk superconductor \cite{Zhang2018,Clark2018,Nayak2021}. These mechanisms generally lead to time-reversal  invariant topological states \cite{Fu2008}, and their critical temperature is determined by that of the bulk superconductor.  An alternative approach, which expands the limits of these techniques, is utilizing flat surface bands which can support superconductivity without a proximity effect. Such superconducting surface states have been observed in KZnBi \cite{Song2021} and t-PtBi$_2$ \cite{Schimmel2023}. However, the properties of these states, particularly the origin and nature of their superconductivity, are not fully understood.

Here, we explore the possible formation of a superconducting surface state at the (111) surface  of the half Heusler compound YPtBi through extensive measurements that include transport, ac susceptibility measurements through mutual inductance, scanning tunneling microscopy and spectroscopy and polar Kerr effect. While robust bulk superconductivity appears below $T_c^b$, we find that superconductivity at the (111) surface persists to temperatures $T_c^s \approx 3T_c^b$.  Furthermore, our Kerr effect measurements suggest that unlike the bulk,  this superconducting surface state breaks time reversal symmetry (TRS), thus revealing  a delicate interplay between the surface and bulk states. We complement these measurements with scanning tunneling spectroscopy and mutual inductance of (111) oriented LuPtBi to substantiate our arguments. To further support our experimental findings, we performed ab-initio calculations on the Bi, Lu and Pt terminations of the (111) direction, finding that a Bi-termination surface exhibits a van-Hove singularity (VHS) near the Fermi energy, which could be a source for for the superconducting surface state. A recent theoretical study \cite{Schwemmer2022} explored a similar surface band structure, finding a possible domain for the onset of chiral topological surface superconductivity ahead of the bulk transition, in agreement with our results.  Our results suggest a new platform for 2D chiral superconductors in which a surface state condenses into a TRS-breaking superconducting state independently from the bulk. 

In a half-Heusler compound, three metallic elements are arranged in interpenetrating fcc lattices ($C1_b$ structure) \cite{Graf2011}. The choice of elements, with thousands of possible combinations, enables control over various properties in the resulting compounds: structures can be metallic, semiconducting, magnetic or non-magnetic, superconducting, and they can have topological surface states \cite{Manna2018,Tavares2023}. Of these materials, YPtBi and LuPtBi have the largest inverted gaps \cite{Chadov2010}, leading to topologically protected surface states \cite{Liu2016}. Furthermore, YPtBi and LuPtBi are bulk superconductors, with critical temperatures of 0.8 K and 1 K respectively \cite{Butch2011,Tafti2013}. The bulk superconductivity should be a combination of odd and even parity states, due to the lack of inversion symmetry \cite{Smidman}.  While much work was done to explore the bulk superconductivity in these materials \cite{Butch2011,Tafti2013,Bay2012,Bay2014,Meinert2016,Pavlosiuk2016,Timm2017,Yang2017,Savary2017,Kim2018,Venderbos2018,Wang2018,Roy2019,Kim2021,Kim2022,Zhou2023,Souza2023}, studies of the surface states and their propensity to unconventional superconductivity have been limited \cite{Schwemmer2022}.
\bigskip

We studied single crystals of LuPtBi and YPtBi that were grown by self-flux solution growth method, with Bi used as the flux medium \cite{Canfield1991,Canfield1992,supp}. The crystal structure and composition were verified using dispersive X-ray analysis, Laue X-ray diffraction and X-ray Photoelectron Spectroscopy \citep{supp}. Transport measurements confirmed the low carrier densities, $n\sim1\times10^{19}$/cm$^3$, and relatively high mobilities, typically $\mu\sim2500$ cm$^2$/V-s at helium temperature. Sharp bulk superconductivity was verified with both resistive transition and two-coils mutual inductance (MI) measurements. The scanning Tunneling Microscopy (STM) and spectroscopy (STS) measurements were performed after removing the top few surface layers with the STM tip itself. The details of this surface cleaning procedure are described in the Supplemental Material \cite{supp}, and are based on a similar procedure previously used to expose superconducting regions in vacuum-cleaved YPtBi crystals \cite{Baek2015}.

\begin{figure}[h]
\includegraphics[width=1.0\columnwidth]{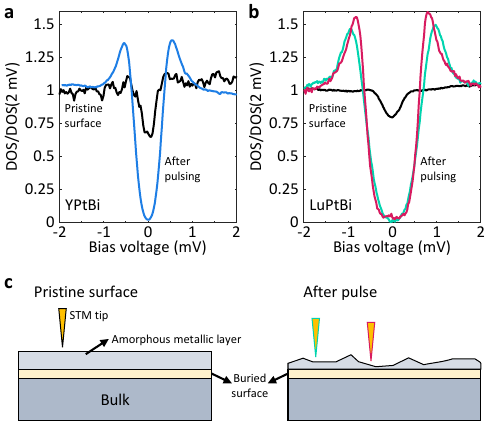}
\caption{Tunneling spectra taken at base temperature before and after tip-cleaning for (a) YPtBi and (b) LuPtBi. The pristine surfaces of both compounds show a gap much larger than the gap expected for the respective $T_c$s. (c) Proposed scenario for the effect of tip-cleaning. The amorphous metallic layer at the surface is partially removed, generating a thinner layer with varying thickness.}
\label{tunn}
\end{figure}

Fig. \ref{tunn} shows representative base temperature ($\sim$350 mK) spectra of pristine surface and tip-cleaned surface for YPtBi and LuPtBi samples. While the tip-cleaned surface improved the spectra, particularly in reducing the observed zero bias conductance to very low values, the pristine surface spectra pointed to a similar large gap. The observed spectra are rounded, even well below temperatures corresponding to the observed large gap. Such a behavior could be because of a nodal gap structure or high effective temperature due to RF heating. The observation of a flat gap (Fig.~\ref{tunn}b) rules these options out. Another possibility is that a thin metallic layer between the superconducting state and the tip depresses the gap by proximity effect (Fig. \ref{tunn}c) \cite{Moussy2001}. In this scenario, the variability in the flatness of the gap and the density of states at zero bias \cite{supp} is explained by the inhomogeneous metallic film that is created by the tip cleaning. Figure \ref{tunn}c illustrates the proposed effect of the current pulse. Before the pulse, a thin disordered metallic layer covers the surface, and the pulse removes some, but not all, of this layer, resulting in a variable metallic barrier, which rounds or supresses the gap at different spots.

\begin{figure}[h]
\includegraphics[width=1.0\columnwidth]{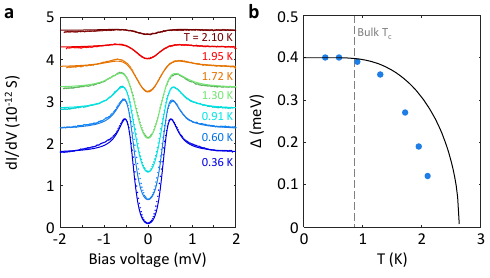}
  \caption{Temperature dependence of tunneling spectra for YPtBi. (a) Spectra take at various temperatures. Solid lines are fits to a Dynes formula (see \cite{supp}). (b) Extracted gap as a function of temperature The solid line is the BCS gap function, assuming $2\Delta/k_BT_c=3.52$.}
	\label{fitsgaps}
\end{figure}

The large gaps observed at base temperature indicate that there is a superconducting system in both LuPtBi and YPtBi which involves a transition temperature higher than the bulk. Indeed, temperature dependence of tunneling spectra of both materials indicates gaps that persist much above the bulk $T_c$. Fig.~\ref{fitsgaps}a shows a typical set of spectra taken on YPtBi at different temperatures. To estimate the gap, we fitted the spectra to the Dynes formula  \cite{Dynes1978,supp}. The temperature was left as a fitting parameter, with a higher effective temperature required to account for the broadening of the spectra. Figure \ref{fitsgaps}b shows the extracted gaps plotted against the measured temperature. The gap at our base temperature was 0.4 meV. Assuming $2\Delta/k_BT_c=3.52$ yields $T_c \approx 2.64$. We note that while the spectra taken following the tip-cleaning procedure can be fitted to the Dynes formula, this procedure fails for the spectra taken on the pristine surface, due to the large residual density of states. This is consistent with the presence of a thin metallic barrier between the tip and the superconducting layer, as has been observed and modeled for thin metal-supercodnuctor interfaces \cite{Moussy2001}. This metallic barrier is then removed by the pulsing procedure.

The tip cleaning procedure could change the composition on the surface or the lattice structure. Indeed, the procedure used in Ref. \cite{Baek2015} generated craters around the pulse location, in which a polycrystalline system was formed. The confinement of superconductivity to these craters could result in an increaed gap. However, the gaps we observed before and after the tip-cleaning procedure were of similar sizes, suggesting that the cleaning does not induce the larger gaps. Furthermore, the topography of our samples following the pulsing procedure did not show craters, and spectra did not systematically vary with the distance from the pulse \cite{supp}. 

\begin{figure}[h]
\includegraphics[width=1.0\columnwidth]{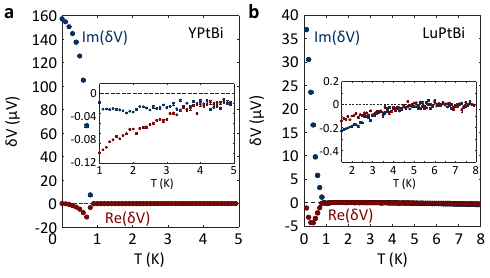}
\caption{Mutual inductance determination of bulk superconducting transition for (a) YPtBi, and (b) LuPtBi. Insets show the onset of a small diamagnetic signal at $\sim$3 K YPtBi. In LuPtBi, a quadratic fit to the high temperature data was removed. This background is due to a thermally-induced change in the distance between the coil and the sample\cite{supp}.}
\label{LYMI}
\end{figure}

A tip-induced enhancement of superconductivity is further ruled out by measurements of the ac susceptibility of both materials, using a mutual inductance technique \cite{Zhang2023}. These crystals were not treated before the measurements. Both materials showed a large increase of the diamagnetic response at their respective bulk transition temperatures (Fig. \ref{LYMI}), in agreement with previous reports \cite{Butch2011,Tafti2013}. However, a careful measurement of YPtBi revealed a small increase, $\sim$0.5\% of the change due to the bulk transition, which onsets at $\sim$3 K (Fig. \ref{LYMI}a). LuPtBi showed a similar change, at around 5 K, over a background that is caused by thermal expansion (Fig. \ref{LYMI}b and \cite{supp}).  This may indicate that a small volume fraction of both samples turns superconducting at a higher temperature than the bulk.

The observation of a similar transition temperature for the vanishing of the superconducting gap in tunneling, and for the onset of an inductive signal in mutual inductance points to an intrinsic origin of superconductivity, that onsets at a temperature much above the bulk $T_c^b$. We envision two possible origins for the observations. First, given our inability to make clean, atomically flat cleaves that are known to expose the (111) plane, the signals may be attributed to another compound, created during the surface cleaning, a combination of the HH elements, possibly with contaminants, particularly oxygen. Superconducting compounds that may be formed this way are listed in Table~\ref{table}. While the $T_c$ of both amorphous and polycrystalline Bi are suitable, such compounds on the surface would generate similar transitions for both LuPtBi and YPtBi, inconsistent with the difference in gap sizes and onset temperatures. We further note that LuPtBi and YPtBi are very similar chemically and structurally (lattice constants $a=6.58$\AA\ and $a=6.65$\AA, respectively \cite{Chadov2010}), suggesting that a strained inclusion of a different compound on the surfaces would yield a similar gap. Moreover, the similarity in the behavior of LuBi and YBi may rule this possibility out as well, particularly since they require pressure to exhibit superconductivity. \begin{table}[ht] 
\begin{tabular}{|c||c|c|c|c|c|c|c|}
\hline
Compound&Bi&a-Bi&p-Bi&BiPt&Bi$_2$Pt&YBi&LuBi \\
\hline
P (GPa)&amb.&amb.&amb.&amb.&5-6&$\gtrsim2.6$&$\gtrsim3$ \\
\hline
$T_c (K)$&$\ll0.1^1$&5-6$^2$&6.1$^3$&1.3$^4$&$\sim$2$^5$&$\sim$5$^6$&4-7.5$^7$ \\
\hline
\end{tabular}
\caption{Possible superconducting compounds derived from YPtBi and LuPtBi. While ``amb.'' represents ambient pressure, some compounds only become superconducting at a finite pressure.  Source references: $^1$\cite{Prakash2017}, $^2$\cite{Chen1967}, $^3$\cite{Petrosyan1973,Tian2006}, $^4$\cite{Zhuravlev1959}, $^5$\cite{Wang2021}, $^6$\cite{Xu2019}, $^7$\cite{Gu2020}.}
\label{table}
%\caption*{References: $^1$\cite{Prakash2017}, $^2$\cite{Chen1967}, $^3$\cite{Petrosyan1973,Tian2006}, $^4$\cite{Zhuravlev1959}, $^5$\cite{Wang2021}, $^6$\cite{Xu2019}, $^7$\cite{Gu2020}.}
\end{table}

The alternative explanation for the observed higher $T_c$ in the two compounds is the creation of a chiral surface state. Particularly, since both LuPtBi and YPtBi are known to be topological with a large band inversion \cite{Chadov2010}, it was suggested that superconductivity associated with this material is topological and may break TRS \cite{Savary2017,Kim2018}. To test this possibility, we performed high resolution polar Kerr effect measurements on YPtBi using a zero-area-loop Sagnac interferometer (ZALSI) \cite{Xia2006,Kapi2009} (see supplementary material \cite{supp}). 
 \begin{figure}[h]
\includegraphics[width=1.0\columnwidth]{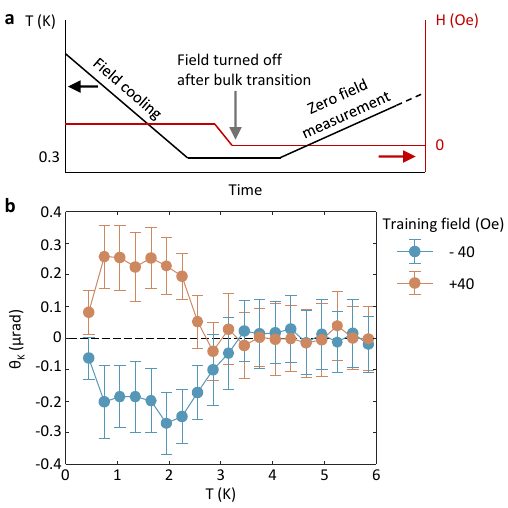}
\caption{Polar Kerr effect measurements on a (111) face of a pristine YPtBi crystal. (a) The sample was first cooled in a magnetic field to base temperature ($\sim$300 mK). The field was then removed, and Kerr data was taken while warming up the sample. (b) The resulting Kerr signals following training in $\pm$40 Oe. A finite Kerr signal appears much above the bulk $T_c^b$ of the sample. The error bars represent a 95\% confidence interval, based on statistical averaging within each temperature bin.}
\label{trsb}
\end{figure}

We cooled down the sample to $\sim$300 mK from a high temperature in the presence of a small external magnetic field, and then removed the field and measured the Kerr signal while warming up the sample (Fig. ~\ref{trsb}a). For small training fields ($H\lesssim50$ Oe) the initial signal (at 300 mK) following the removal of the magnetic field was very small, consistent with zero (within our resolution of $\sim0.05 \ \mu$rad). However, increasing the temperature from base temperature, a finite Kerr signal of order $\pm0.25 \ \mu$rad appeared above the bulk $T_c^b$. The signal was independent of the field strength and changed sign upon changing the direction of the training field (Fig. \ref{trsb}b and \cite{supp}). These features indicate a TRS breaking state, and are generally similar to those observed in chiral superconductors \cite{Schemm2014,Hayes2021}. Note that a TRSB state further restricts the association of the signals observed using mutual inductance and STM with a different chemical structure of the surface. In particular, amorphous and polycrystalline Bi are not chiral superconductors, and therefore will not give rise to a finite Kerr effect. The contribution from vortices \cite{supp} trapped by the field-cooling procedure is too small to account for the observed signals.

With strong structural and electronic structure similarities to GdPtBi, we estimate the optical penetration depth of YPtBi at $\sim 150$ nm \cite{Hutt2018}. Thus, the Kerr signal can have both bulk and surface contributions.  In particular above the bulk $T_c^b$ the observed Kerr signal may be a result of TRS breaking associated with the surface superconductivity - a favorable possibility since Kerr effect persists very close to the surface $T_c^s$ , or it may be associated with vestigial order \cite{Bojesen2013,Bojesen2014,Fernandes2019,How2023} originating from the bulk superconducting state. However, the suppression of the Kerr rotation below $T_c^b$, as well as the nodes in the gap and analysis of the possible spin and orbital pairing states \cite{Kim2018} suggest that the bulk does not break TRS.

\begin{figure}[h]
\includegraphics[width=1.0\columnwidth]{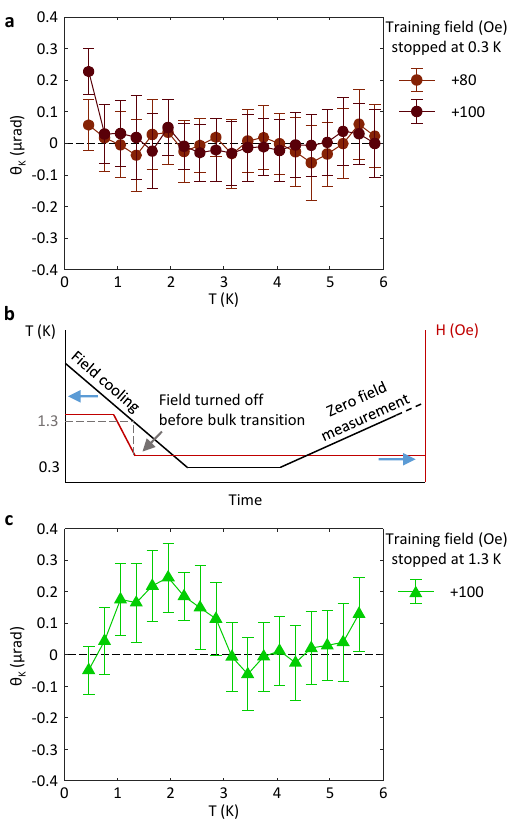}
\caption{Polar Kerr effect after higher-field training. (a) Absence of Kerr rotation measured while warming up after training at high fields ($>$ 80 Oe). (b) An alternative training protocol, where the magnetic field was removed at 1.3 K, before the bulk superconducting transition. (c) The Kerr rotation following the protocol in b is recovered.}
\label{trsb_HF}
\end{figure}

While low field training (i.e. training in $H\lesssim50$ Oe) yields similar onset and size of Kerr signal, which follows the direction of the training field, cooling the sample at higher magnetic field to base temperature (0.3 K), turning the field to zero and measuring while warming up the sample in zero magnetic field yields a signal that is zero within our resolution of $\sim0.05 \ \mu$rad (Fig. \ref{trsb_HF}a). However, cooling the sample at higher magnetic field to base temperature (0.3 K), but on the way turning the magnetic field to zero at $T\approx 1.3$ K, then measuring the Kerr effect at zero field from base temperature up yields results similar to the low field training experiments (Fig.~\ref{trsb_HF}b,c).

Such a peculiar effect may point to a delicate interplay between the vortex state that appears in the bulk material upon cooling in a field much below the bulk $T_c^b$, and the TRSB surface state. For example, cooling at lower field and removing the field at low temperatures may result in low vortex occupancy due to weak surface and geometrical barriers (note that $H_{c1}\lesssim3$ Oe, while the thermodynamic critical field is $\lesssim100$ Oe for the bulk superconducting state of these materials). Above a field of order $\gtrsim100$ Oe, enough vortices may remain in the bulk, which upon crossing the bulk Tc will move out of the sample, thus possibly  ``scrambling'' the single domain nature of the TRS-breaking state of the surface superconductivity. Thus, the magnetization of the surface state interacts with a conventional bulk superconducting order parameter. This is similar to magnetic superconductors, where the formation and motion of vortices can change the ferromagnetic order parameter \cite{Paulsen2012}.

Taken together, the STM, mutual inductance and PKE results indicate that the (111) surfaces of YPtBi and LuPtBi host a superconudcting state which breaks TRS. To understand the origin of this state, we studied the surface band structure of LuPtBi (see Supplementary Information \cite{supp}). In Bi-terminated (111) surfaces, we found a van-Hove signularity close to the $M$-point, resulting in an enhanced density of states near the Fermi level, which could give rise to the enhanced surface $T_c^s$. The calculated band structure of the Lu terminated surface did not show van-Hove singularities. Schwemmer {\it et al.} \cite{Schwemmer2022} have recently reported a detailed study of the surface band structure and possible superconductivity of the Bi-terminated (111) surface of LuPtBi. While the gross features of their surface band structure are similar to our calculation, they also revealed flat-band regions near the Fermi energy, and calculated the propensity to a superconducting instability. In particular, they argue that the transition temperature of the surface state is independent of the bulk transition, and exhibit vestigial time-reversal symmetry breaking in it fluctuation regime.  Since inversion symmetry is broken, simple singlet and/or triplet states were dismissed and pairing was tested for pairs of electrons that are connected by the TR operator. The result has been a pair-amplitude that transforms under the E irreducible representation of the surface $C_{3v}$ point group (relevant to the (111) crystal orientation) in the presence of strong Kane-Mele type and Rashba type spin-orbit coupling \cite{Laubach2014}. Furthermore, they found that a chiral superconducting state, which spontaneously breaks TRS, is energentically favorable. While we focused on the (111) oriented surface, surface bands close to the Fermi energy that originate from dangling bonds are generic in the half-Heuslers along other surfaces, including the (001) \cite{Liu2016,Kawasaki2018} and (120) \cite{Souza2023}. It is therefore possible that the surface superconductivity we report here could also exist on other surfaces, depending on the filling and flatness of the surface bands.

In conclusion, we presented a comprehensive study of the (111) surface of the half-Heusler alloys YPtBi and LuPtBi. While a bulk $T_c^b$ of $\sim$0.8 K is observed for both systems, we find a much larger corresponding gap in our tunneling studies with $T_c^s$'s of $\sim$2.3 K and $\sim$6.5 K for YPtBi and LuPtBi, respectively. Mutual inductance measurements of YPtBi further revealed a small diamagnetic response that onsets around $T_c^s$. Polar Kerr effect on YPtBi indicates that TRS is broken near or at $T_c^s$. Together, our data suggest that chiral surface superconductivity precedes bulk condensation in these materials, a hypothesis which is strongly supported by recent calculations \cite{Schwemmer2022}. While additional work is required to determine the order parameter of these states, our results open a new direction to engineering topological superconducting states in half Heusler compounds. Although our study was conducted on single crystals, Kim {\it et al.} \cite{Kim2023} recently demonstrated epitaxially growth of YPtBi films, which can also be encapsulated by an alumina film, preventing degradation of the surface. This system is therefore highly promising for quantum device applications.

\acknowledgments
We thank Ronny Thomale and J\"org Schmalian for illuminating discussions. This work was supported by  the U.~S.~Department of Energy (DOE) Office of Basic Energy Science, Division of Materials Science and Engineering at Stanford under contract No.~DE-AC02-76SF00515 and by the National Science Foundation Award Number: 2307132. EP was partially supported by the Koret Foundation. 

\end{document}

\title{SUPPLEMENTAL INFORMATION FOR\\Possible Unconventional Surface Superconductivity in the Half-Heusler YPtBi}

\author{Eylon Persky}
\email{perskye1@stanford.edu}
\affiliation{Geballe Laboratory for Advanced Materials, Stanford University, Stanford, CA 94305, USA}
\affiliation{Stanford Institute for Materials and Energy Sciences, SLAC National Accelerator Laboratory, 2575 Sand Hill Road, Menlo Park, CA 94025, USA}
\affiliation{Department of Applied Physics, Stanford University, Stanford, CA 94305, USA}
\author{Alan Fang}
\affiliation{Geballe Laboratory for Advanced Materials, Stanford University, Stanford, CA 94305, USA}
\affiliation{Stanford Institute for Materials and Energy Sciences, SLAC National Accelerator Laboratory, 2575 Sand Hill Road, Menlo Park, CA 94025, USA}
\author{Xinyang Zhang}
\affiliation{Geballe Laboratory for Advanced Materials, Stanford University, Stanford, CA 94305, USA}
\affiliation{Stanford Institute for Materials and Energy Sciences, SLAC National Accelerator Laboratory, 2575 Sand Hill Road, Menlo Park, CA 94025, USA}
\affiliation{Department of Applied Physics, Stanford University, Stanford, CA 94305, USA}
\author{Carolina Adamo}
\altaffiliation[Current address: ]{Northrop Grumman Corporation, Redondo Beach, CA 90278, USA}
\affiliation{Geballe Laboratory for Advanced Materials, Stanford University, Stanford, CA 94305, USA}
\affiliation{Stanford Institute for Materials and Energy Sciences, SLAC National Accelerator Laboratory, 2575 Sand Hill Road, Menlo Park, CA 94025, USA}
\author{Phillip Wu}
\altaffiliation[Current address: ]{Center of Science Education, National Chung-Hsing University, Department of Chemistry, Taichung City 402, Taiwan ROC}
\affiliation{Geballe Laboratory for Advanced Materials, Stanford University, Stanford, CA 94305, USA}
\author{Eli Levenson-Falk}
\altaffiliation[Current address: ]{Center for Quantum Information Science and Technology, University of Southern California, Los Angeles, California 90089, USA}
\affiliation{Geballe Laboratory for Advanced Materials, Stanford University, Stanford, CA 94305, USA}
\affiliation{Department of Applied Physics, Stanford University, Stanford, CA 94305, USA}
\author{Chandra Shekhar}
\affiliation{Max Planck Institute for Chemical Physics of Solids, 01187 Dresden, Germany.}
\author{Claudia Felser}
\affiliation{Max Planck Institute for Chemical Physics of Solids, 01187 Dresden, Germany.}
\author{Binghai Yan}
\affiliation{Department of Condensed Matter Physics, Weizmann Institute of Science, Rehovot, Israel}
\author{Aharon Kapitulnik}
\email{aharonk@stanford.edu}
\affiliation{Geballe Laboratory for Advanced Materials, Stanford University, Stanford, CA 94305, USA}
\affiliation{Stanford Institute for Materials and Energy Sciences, SLAC National Accelerator Laboratory, 2575 Sand Hill Road, Menlo Park, CA 94025, USA}
\affiliation{Department of Applied Physics, Stanford University, Stanford, CA 94305, USA}
\affiliation{Department of Physics, Stanford University, Stanford, CA 94305, USA}

\date{\today}
\maketitle

\onecolumngrid
\newpage

\renewcommand{\thefigure}{S\arabic{figure}}

\section{Crystal growth and sample fabrication}

Single crystals of YPtBi and LuPtBi were grown by self-flux solution growth method, whereas Bi provides a flux medium. The stoichiometric quantities of freshly polished piece of elements Lu, Pt and Bi of purity $>99.99 \%$ in the 0.3:0.3:10 atomic ratio were put in tantalum crucible and sealed in dry quartz ampoule under 3 mbar partial pressure of argon. The mixture-containing ampoule was heated at a rate of 100 K/h to 1473 K, followed by 12 hours soaking. For crystal growth, the temperature was slowly reduced by 2 K/h to 873 K, with extra Bi flux removed by decanting the ampoule at 873 K. Using this method, we obtained 2-3 mm regular triangle size of crystals. The preferred growth orientation was (111), which was confirmed by Laue diffraction. The method of crystal growth has been adapted from \cite{Canfield1991,Canfield1992}. The composition and structure were checked by energy dispersive X-ray analysis and Laue X-ray diffraction, respectively. The lattice parameters of the cubic structure are 6.574\AA, consistent with previous reports \cite{Chadov2010,Tafti2013}.  Energy dispersive x-ray spectrometry gives atomic percentages 33.6: 33.0: 33.4 $\pm$ 3.0 $\%$ for Lu: Pt: Bi, confirming the stoichiometric ratio of the chemical compositions. Similar procedure and results were obtained for YPtBi.

\section{Surface Properties and Morphology}

\begin{figure}[h]
\begin{varwidth}{\linewidth}
\includegraphics[width=1\textwidth]{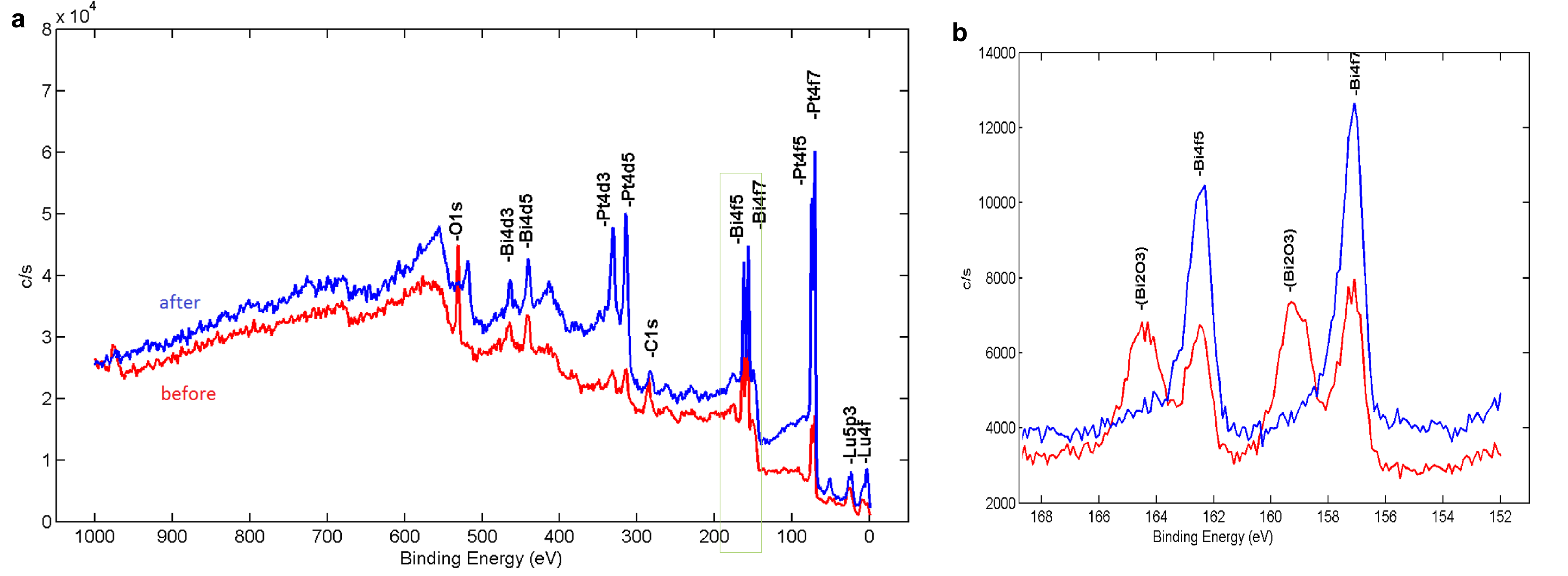}
\caption{XPS studies of a LuPtBi crystal before and after ion milling. We also show an expanded region around 160 eV, showing the removal of bismuth-oxide and strengthening the LuPtBi-bonded bismuth.}
\label{fig-milling}
\end{varwidth} 
\end{figure}

 \begin{figure}[h]
 \begin{varwidth}{\linewidth}
\includegraphics[width=1\textwidth]{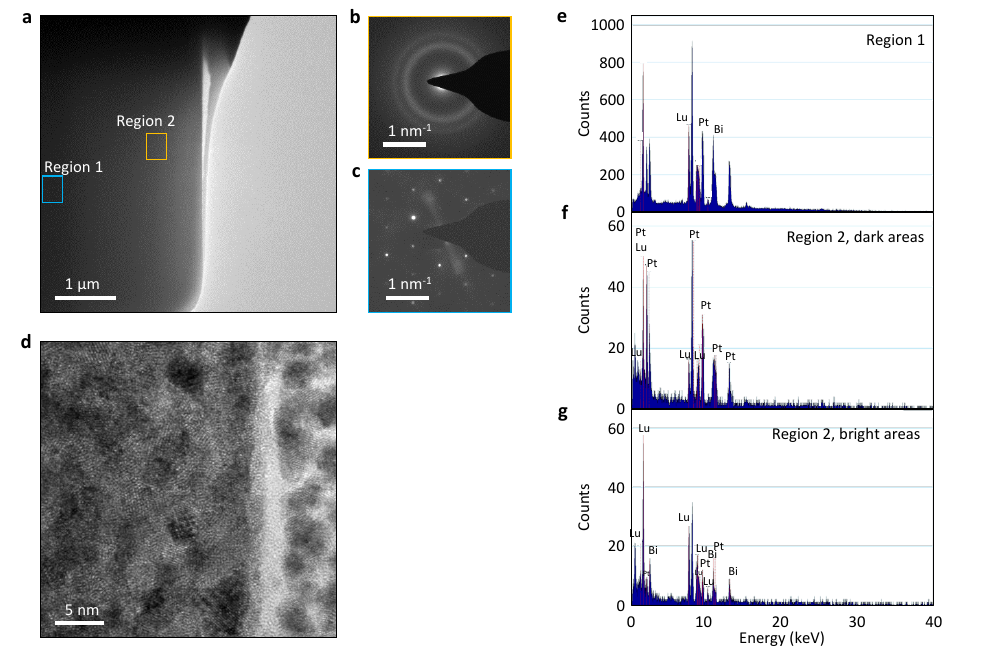}
\caption{TEM and EDS characterization of the pristine surface. (a) TEM cross section showing a single crystal in the bulk of the material, and a polycristalline region near the surface. (b,c) Diffraction pattern in the regions marked 1 and 2, showing a crystalline structure away from the surface. (d) A high resolution image of the surface of the sample, showing two different types of polycrystalline regions, that appear as dark and bright regions.  (e-g) EDS spectra taken on the bottom part of the crystal (e), and the darker (f) and brighter polycrystalites (g). In the single-crystal region, Lu, Pt and Bi appear in equal ratios, and there is no significant oxygen peak, whereas the dark and bright areas correspond to PtO-rich and LuBiO-rich regions.}
\end{varwidth} 
\label{TEM}
\end{figure}

To understand the morphology and chemistry of the surface of the crystals, we used X-ray Photoelectron Spectroscopy (XPS) that allows for high sensitivity elemental surface composition measurement and high resolution binding energy chemical shift measurement of solid samples under high vacuum (Fig. \ref{fig-milling}).  With typical measurement depth of about a nanometer, pristine crystals show traces of carbon, oxygen, and bismuth on the surface. However, after a several seconds of ion milling (equivalent to a few nm), all carbon traces are gone and all three elements: Y/Lu, Pt and Bi seem to appear with the correct 1:1:1 ratio, together with traces of oxygen. This did not change after subsequent milling of several more nanometers of the surface, confirming that no other elements exist in the samples. Figure \ref{TEM} shows transmission electron microscope (TEM) data of a cross section of a LuPtBi sample.The images reveal that in the regions close to the surface of the sample, the structure consists of polycrystalline patches that are several nanometers in size. Electron diffraction spectroscopy (EDS) showed equal magnitudes of the Lu, Pt abd Bi peaks in the spectra in the bulk, but the polycrystalline regions were PtO-rich or LuBiO rich.

\section{Surface cleaning procedure using STM tip}
\begin{figure}[h]
\includegraphics[width=0.5\columnwidth]{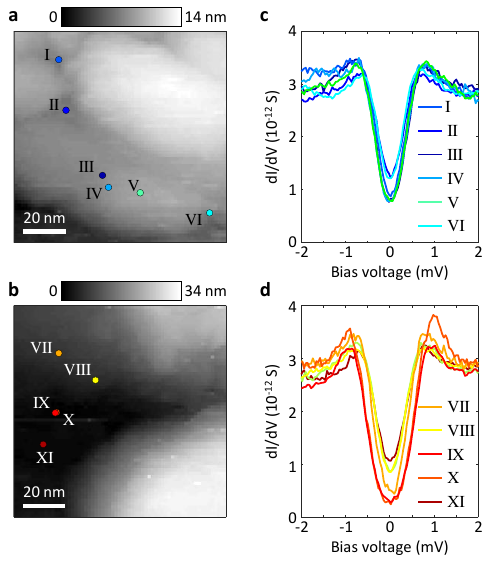}
	\caption{(a,b) Surface topography measured after the tip cleaning procedure, at three locations on a LuPtBi sample. (c,d) Corresponding spectra taken on various points on the cleaned surfaces. Both the topography and the spectra do not systematically vary with the distance from the pulse location.}
	\label{Topo}
\end{figure}

The ease by which the fresh surface oxidizes suggests that ex-situ milling and then loading the sample into the STM may not solve the surface contamination problem. To overcome this issue, we performed in-situ surface cleaning by using the STM tip to remove the top few layers and expose deeper layers. This process was performed in-situ at helium temperature under UHV conditions. The STM tip junction resistance was set by a bias of 20-50 mV and current of 100 pA. Then voltage pulses of a few volts were applied until the zero bias conductance was minimized. Occasionally, we found regions of area of $\sim 5\times 5$ nm$^2$, which were flat and untilted to within a few angstroms,  yet had an indistinct but reproducible topography (no atomic corrugations above 10's of pm were seen, most likely due to tip bluntness). Since we focused on the temperature range below $T_c^b$, the quality of spectra were gauged by how close the DOS reached zero at zero bias, as expected for gapped spectra. A similar method was previously used to expose superconducting regions in vacuum-cleaved YPtBi crystals by Baek {\it et al.} \cite{Baek2015}, although no particular effort was made to study the (111) direction of the crystals. Notably, our (111) starting surface, while showing ``lumpy'' topography similar to \cite{Baek2015}, was overall flatter, with only $\sim$0.3 nm height variations over 50 nm regions (as opposed to .$\sim$1 nm in \cite{Baek2015}). 
Fig. \ref{Topo}a,b shows the topography of a LuPtBi crystal after tip cleaning at two different locations. Unlike Ref. \cite{Baek2015}, we do not observe a "crater" topography around the cleaned region. Furthermore, spectra measured at various locations after the cleaning (Figure \ref{Topo}c,d) do systematically depend on the position. 

For the mutual inductance and polar Kerr effect measurements, data were taken on pristine surfaces without any ex-situ preparation.

\section{STM Measurements}
\begin{figure}[h]
\includegraphics[width=1\columnwidth]{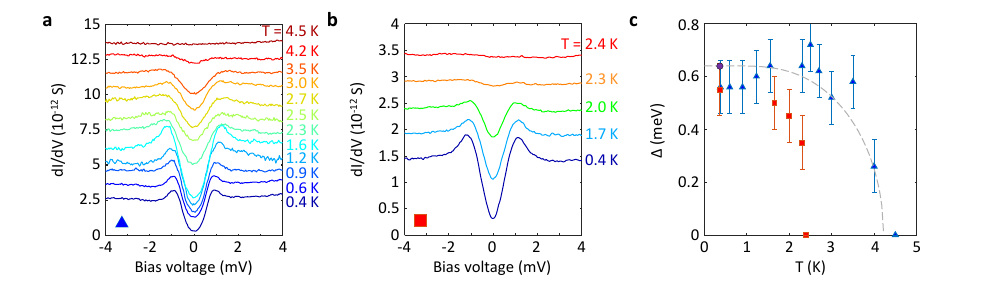}
	\caption{ (a,b) Examples of surface spectroscopy following tip-cleaning procedure as a function of temperature on LuPtBi. (c) Superconducting gaps as a function of temperature, estimated from the spectra. The dashed line is a BCS gap function, with a gap of 0.64 meV, as seen in main Fig. 1b (purple dot). The error bars represent the systematic error from estimating the gap from the maximum slope of the spectra.}
	\label{LYT}
\end{figure}

\begin{figure}[h]
\includegraphics[width=0.5\columnwidth]{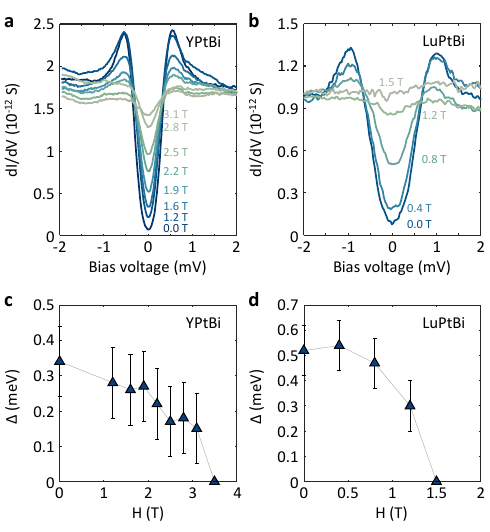}
	\caption{ Examples of surface spectroscopy following tip-cleaning procedure performed at base temperature (365 mK) as a function of magnetic field  for a) YPtBi and b) LuPtBi. The gaps as a function of field shown in (c,d) were estimated using the maximum slope of the spectra. }	
	\label{LYH}
\end{figure}

STM measurements were preformed on a modified UNISOKU Co., USM1300S$^3$He\textsuperscript{TM} system with Ultra High Vacuum (UHV), 350 mK base temperature, and magnetic field capabilities. Figs. \ref{LYT} and Fig. \ref{LYH} show additional tunneling spectra as a function of temperature and magnetic field, for both LuPtBi and YPtBi.

To obtain the gap, the spectra were fitted to the Dynes formula
\begin{equation}
    \frac{dI}{dV} \propto \int dE \frac{\partial f}{\partial E} \rho (E-V),
\end{equation}
where $f(E)$ is the Fermi-Dirac function and
\begin{equation}
    \rho(E) = \frac{E-i\Gamma}{\sqrt{(E-i\Gamma)^2 - \Delta^2}}
\end{equation}
is the density of states, where $\Gamma$ is the Dynes parameter, which was set to 2 $\mu$eV, and $\Delta$ is the superconducting gap. The temperature (entering into the Fermi-Dirac function) was used as a fitting parameter in order to account for a higher electron temperature in the tip, due to RF radiation. For YPtBi, We found that the effective temperature was $ T_{\text{eff}} = 0.5 \text{\, K} + T_{\text{sample}}$. 

In some cases, the DOS after the cleaning procedure still had residual density of states at zero bias. In these cases, we estimated the gaps from the maximum slope of the spectra \cite{Kita1995}. Figure \ref{LYT} shows spectra obtained on cleaned LuPtBi at two locations, as a function of temperature, and the extracted gaps. While the gaps appear to close at different temperatures, the base temperature spectra have similar gaps.
%\begin{figure}[h]
%\includegraphics[width=0.8\columnwidth]{LYvsT.pdf}
%	\caption{Additional spectra taken on the pristine surface of YPtBi and LuPtBi.}
%	\label{PristineSpectra}
%\end{figure}
 
\bigskip

\bigskip
\section{Mutual inductance measurements}
Mutual inductance measurements were conducted with a gradiometric-type setup, where a drive coil and counter-wound pickup coils are positioned above the sample. A small magnetic field was applied by driving an ac current (3 MHz) through the drive coil, and the cpmplex ac response from the sample was detected using the pick-up coils. The relative phase was set so that at the lowest temperature, the real part of the response was zero. We binned the data into 0.1 K bins and present the mean value in each bin along with an error bar representing the 95\% confidence interval.

\begin{figure}[h]
\includegraphics[width=0.5\columnwidth]{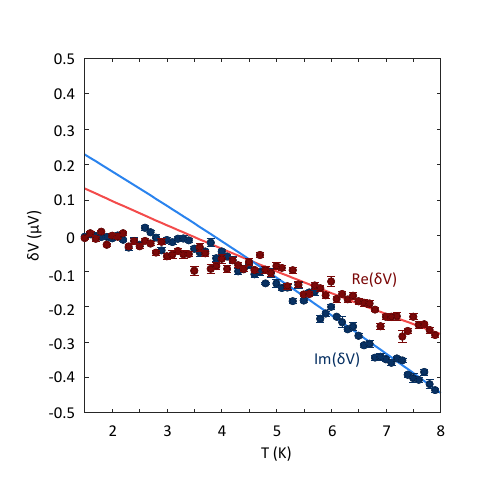}
	\caption{Raw mutual inductance data for LuPtBi. The solid lines are quadratic fits to the data above 5 K, which were subtracted from Fig. 3b of the main text.}
	\label{LPT_MI_raw}
    \end{figure}

For LuPtBi, the data include temperature variations that are due to thermal expansion, which changes the distance between the coil arrangement and the crystal. Fig. \ref{LPT_MI_raw} shows the raw data above the bulk superconducting transitions, along with a quadratic fit to the high temperature portion of the data (above 5 K). There is a clear deviation from the quadratic behavior, that onsets below approximatley 4 K.

    \begin{figure}[h]
\includegraphics[width=0.5\columnwidth]{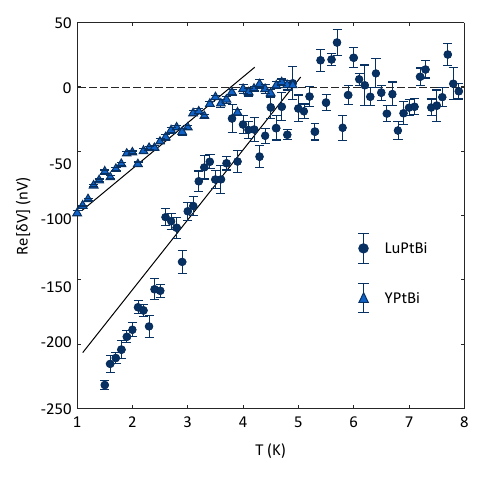}
	\caption{Comparison of the real part of the mutual inductance signal for YPtBi and LuPtBi. The black lines are guides to the eye.}
	\label{Y_Lu_comp}
\end{figure}

\begin{figure}[h]
\includegraphics[width=1\columnwidth]{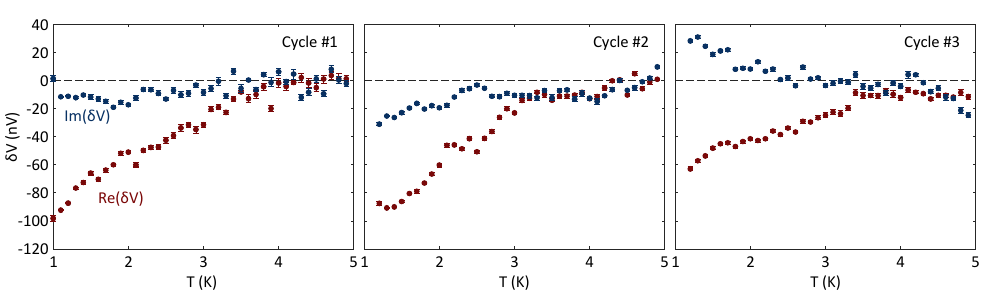}
	\caption{Mutual inductance measurements of the surface transition of YPtBi from subsequent thermal cycles.}
	\label{YPT_MI_repeat}
\end{figure}

\section{Kerr Effect Measurements}
\begin{figure}[h]
\includegraphics[width=1\columnwidth]{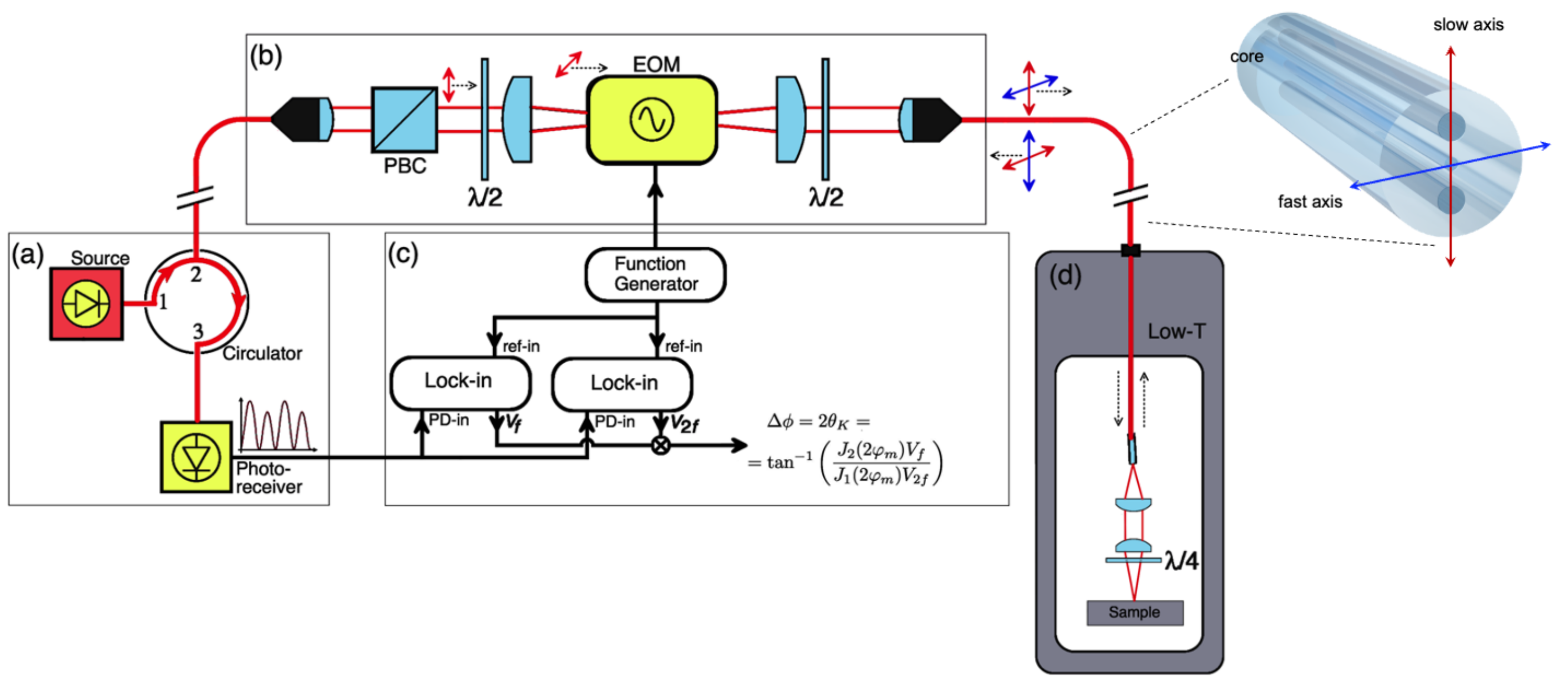}
	\caption{ Schematic of the zero-area loop Sagnac interferometer used in this study. On the right is the end-part showing one beam that enters the quarter waveplate from the fast axis as right circularly polarized, and reflected back into the slow axis as left circularly polarized after experiencing a Kerr angle shift.  A second beam propagates in the direction of the sample in the slow axis and is reflected into the fast axis.}
	\label{zasi}
\end{figure}

High-resolution measurements of Kerr rotation were performed using a fiber-based zero-area loop Sagnac interferometer as first described in Ref.~\cite{Xia2006}. Following the main components shown in Fig.~\ref{zasi}, two linearly polarized counterpropagating beams comprising the interferometer are isolated along the fast and slow axes of $\sim$10 m of polarization-maintaining fiber and are modulated in phase with a frequency that matches the transit time of light through the interferometer. The linearly polarized light traveling along each of these paths passes through a quarter-waveplate and is converted into circularly polarized light just above the sample.  An objective lens focuses the light onto a small interaction region on the sample, and the reflected light beams are converted back to linear polarizations with exchanged polarization directions. In the absence of magnetic field, the apparatus is completely reciprocal by symmetry except for the sample. Thus, upon reflection, one branch of the interferometer acquires a phase shift of $+\theta_K$, while its orthogonally polarized counterpart acquires an opposite phase shift of $-\theta_K$. The two phase shifts are added at the detector and extracted from the ratio of the first and second harmonics of the modulated signal. 

\begin{figure}[h]
\includegraphics[width=1\columnwidth]{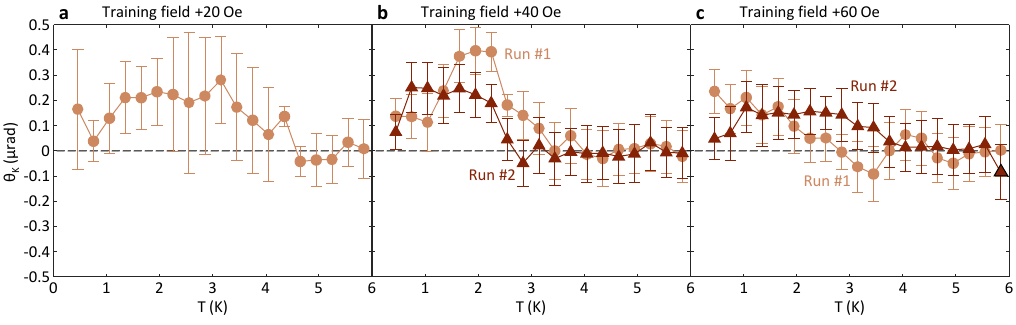}
	\caption{Additional Kerr rotation data taken after training the sample in +20 Oe (a), +40 Oe (b) and +60 Oe (c).}
	\label{LowFieldS}
\end{figure}

\begin{figure}[h]
\includegraphics[width=1\columnwidth]{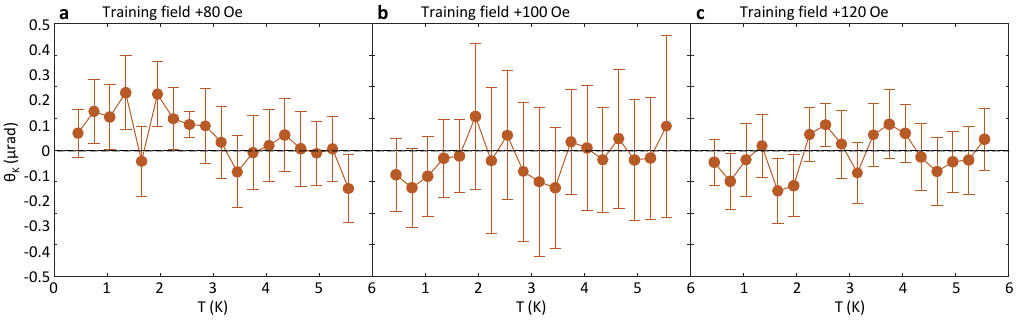}
	\caption{Additional Kerr rotation data taken after training the sample in +80 Oe (a), +100 Oe (b) and +120 Oe (c).}
	\label{HighFieldS}
\end{figure}

\begin{figure}[h]
\includegraphics[width=0.5\columnwidth]{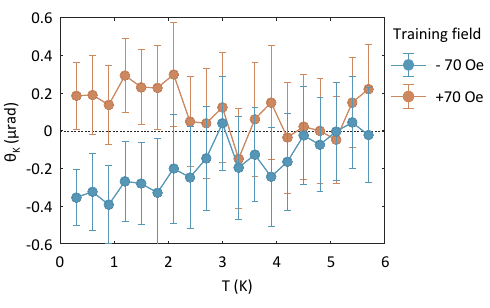}
	\caption{Kerr rotation on a second YPtBi sample.}
	\label{Didi}
\end{figure}

Fig.\ref{LowFieldS} and \ref{HighFieldS} show additional PKE data on YPtBi with various training fields and Fig.\ref{Didi} shows the PKE from a second YPtBi sample.

\bigskip


\subsection{Estimation of vortex contribution to the Kerr signal}
In the Kerr measurements reported here, vortices are trapped in the bulk or surface superconductors due to the training protocol, in which the superconductor is cooled in the presence of a magnetic field. These vortices may be pinned after the field is removed, and therefore contribute to the Kerr rotation measured during the warmup. Below, we provide an estimate for the Kerr rotation due to vortices, and show it is several orders of magnitude lower than the signals observed in the experiment.

In a non-magnetic superconductor, vortices contribute to the Kerr signal via the paramagnetic response of the electrons in the normal core to the field trapped by the vortices. The coherence length of YPtBi is about 15 nm \cite{Kim2021}. Therefore, the magnetic field inside the vortex is $\Phi_0/A_{\text{core}}\approx 2.9$ T, where $\Phi_0$ is the flux quantum and $A_{\text{core}} = \pi\xi^2$ is the area of the vortex core. Assuming a Verdet constant of 1 rad/T/m, and an optical penetration depth of 100 nm \cite{Hutt2018}, the resulting Kerr rotation from a single vortex is $2.9\times10^{-7}$ rad. However, in the experiment, the measured signal is an average of the normal cores and superconducting areas that are within the beam. With a training field $B$, the vortex density is approximately $n_v = B/\Phi_0$ and therefore the normal area fraction is $A_{\text{core}}n_v$. For a training field of 50 Oe, this fraction is 0.2\%, and therefore the contribution from vortices is about 0.6 nrad, which is 1000 times smaller than the measured signal.

\section{Ab-initio calculations of the surface band structure of \LPB}
Ab initio calculations were performed within the density-funcitonal theory framework with the generalized gradient approximation, which is implemented into the VASP package \cite{Kresse1993}. A 54-atomic-layer-thick slab model was employed to simulate the LuPtBi surface, in which the top surface is terminated by Bi and the bottom surface is by Lu. A vacuum layer with more than 50 \AA\, thick is employed in the $z$-direction to avoid inter-slab interactions. A grid of 12$\times$12 is used for the 2D Brillouin zone integration. Spin-orbit coupling was included in all calculations.

\begin{figure}[h]
\includegraphics[width=0.5\columnwidth]{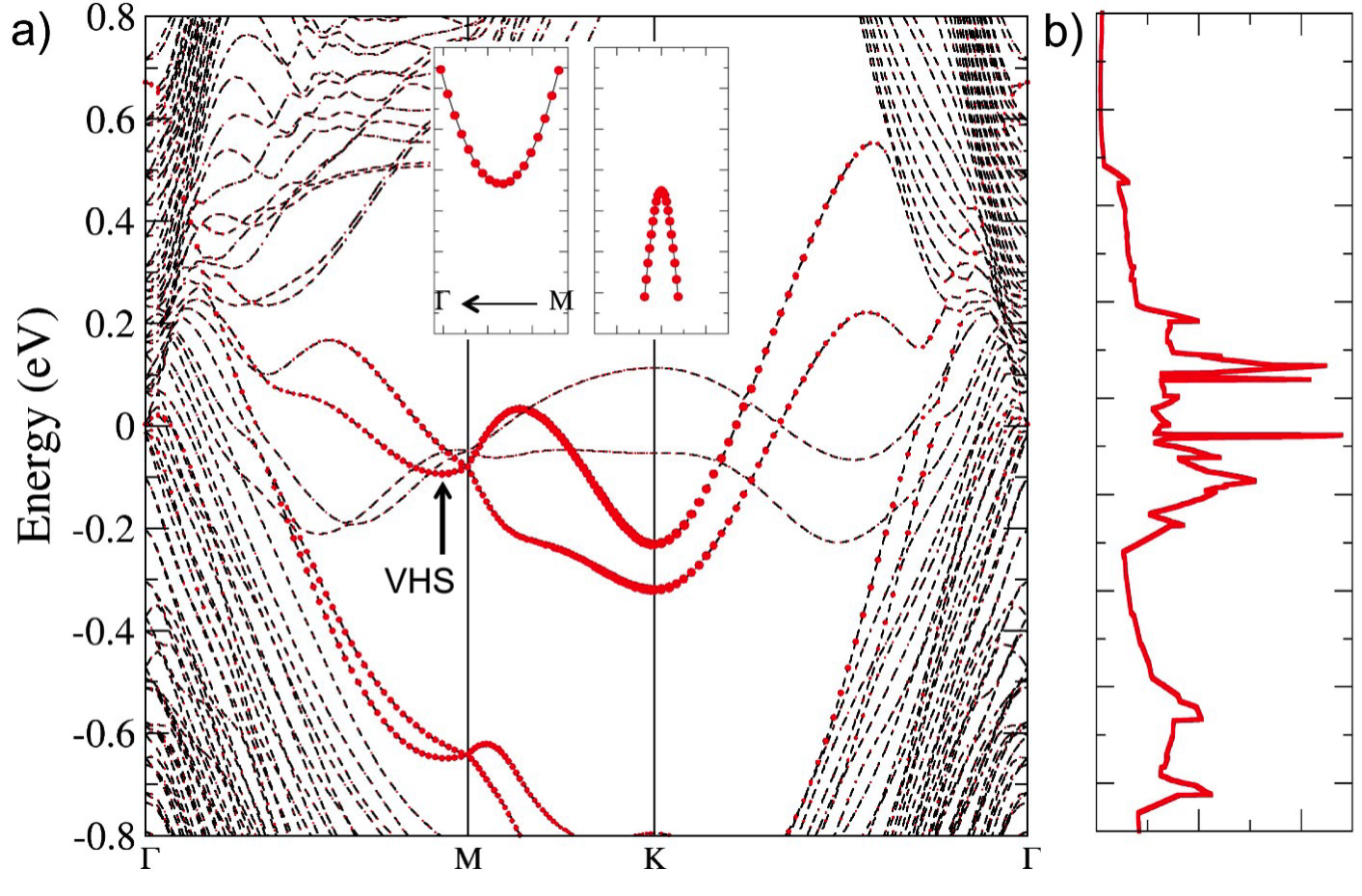}
\caption{The surface band structure and density of states (DOS) from ab-initio calculations. (a) Band structure. The size of red circles represents the projection to the surface Bi atom on the top surface. (b) DOS projected to the surface Bi states. The Fermi energy is shifted to zero. A Van Hove singularity (VHS) point exists in the surface band near the $M$ point (nearly 10\% position from M to $\Gamma$), corresponding to DOS singularity at about $-0.1$ eV. Two insets are shown to demonstrate the VHS point, in which the left inset shows the up-opening parabolic  dispersion along the $\Gamma-M$ line while the right inset shows the down-opening parabolic dispersion along the line perpendicular to the $\Gamma-M$ direction. The saddle (VHS) point is highlighted by blue arrows.}
\label{LPBBS}
\end{figure}

In particular, the Bi-termination surface is shown in Fig.~\ref{LPBBS}, exhibiting a van-Hove singularity (VHS) at $E=-0.10$ eV near the M-point, about 5\% offset along the line from M to $\Gamma$. 6 such VHS points appear in the first Brillouin zone.
